\documentclass[10pt]{article}
\usepackage{fullpage}
\usepackage{amsmath}
\usepackage{amssymb}
\usepackage{amsfonts}
\usepackage{comment}
\usepackage{color}
\usepackage{cite}
\usepackage{subcaption}
\usepackage{graphicx}
\usepackage{enumitem}

\def\##1{{\bf #1}}
\def\=#1{\underline{\underline{#1}}}

\def\+
#1{\underline{\bf #1}}
\def\*#1{\underline{\underline{\bf #1}}}

\def\r#1{(\ref{#1})}
\def\l#1{\label{#1}}

\def\le{\left(}
\def\ri{\right)}
\def\les{\left[}
\def\ris{\right]}
\def\lec{\left\{}
\def\ric{\right\}}
\def\lek{[{\kern 0.1em}}
\def\rik{{\kern 0.1em}]}

\def\.{\mbox{ \tiny{$^\bullet$} }}

\def\eps{\varepsilon}
\def\epso{\eps_{\scriptscriptstyle 0}}

\def\muo{\mu_{\scriptscriptstyle 0}}

\def\ko{k_{\scriptscriptstyle 0}}

\def\ux{\hat{\#u}_{\rm x}}
\def\uy{\hat{\#u}_{\rm y}}
\def\uz{\hat{\#u}_{\rm z}}

\def\inc{_{\rm in}}
\def\host{_{\rm h}}

\def\finc{f\inc}
\def\Si{S\inc}
\def\Vi{V\inc}
\def\vi{v\inc}
\def\Vh{V\host}
\def\Vs{V_{\rm s}}

\def\nh{n\host}

\def\epsi{\eps\inc}
\def\epsh{\eps\host}

\def\Vsp{V_{\rm all}}
\def\Ve{V_{\ell}}

\def\rp{\#r^\prime}
\def\ro{\#r_{\rm o}}

\def\Lt{L_{\rm t}}
\def\Lx{L_{\rm x}}
\def\Ly{L_{\rm y}}
\def\Lz{L_{\rm z}}
\def\epsMGt{\eps^{MG}_{\rm t}}
\def\epsMGx{\eps^{MG}_{\rm x}}
\def\epsMGy{\eps^{MG}_{\rm y}}
\def\epsMGz{\eps^{MG}_{\rm z}}

\begin{document}

\begin{center}

\LARGE{ {\bf Anisotropic homogenized composite mediums arising from  truncated spheres, spheroids, and ellipsoids
}}
\end{center}
\begin{center}
\vspace{10mm} \large
 
 {Tom G. Mackay}\footnote{E--mail: T.Mackay@ed.ac.uk.}\\
{\em School of Mathematics and
   Maxwell Institute for Mathematical Sciences\\
University of Edinburgh, Edinburgh EH9 3FD, UK}\\
and\\
 {\em NanoMM~---~Nanoengineered Metamaterials Group\\ Department of Engineering Science and Mechanics\\
The Pennsylvania State University, University Park, PA 16802--6812,
USA}
 \vspace{3mm}\\
 {Akhlesh  Lakhtakia}\\
 {\em NanoMM~---~Nanoengineered Metamaterials Group\\ Department of Engineering Science and Mechanics\\
The Pennsylvania State University, University Park, PA 16802--6812, USA}

\normalsize

\end{center}

\begin{center}
\vspace{5mm} {\bf Abstract}
\end{center}

Closed-form expressions were recently derived for depolarization dyadics    for
  truncated spheres and truncated spheroids, and the formalism was extended to truncated ellipsoids. These results were exploited to develop an implementation of the 
  Maxwell Garnett homogenization formalism for the relative permittivity parameters of homogenized composite mediums (HCMs) arising from an isotropic host medium impregnated with isotropic 
  inclusions that are truncated spheres, spheroids, and ellipsoids. In so doing, the anisotropy of the HCM was related to the geometry of the inclusions: in general, the more  the shape of the inclusions deviated from spherical, the greater was the degree of anisotropy  exhibited by the HCM.
 
 \vspace{5mm}
\vspace{5mm}

\section{Introduction}

A host medium impregnated with particulate inclusions may be regarded as a homogenized composite medium (HCM) provided that the inclusions are much smaller than the wavelengths involved \cite{SelPap,Neelakanta,Choy,MAEH}. Current interest in  HCMs is  intense
due to the  burgeoning  development of nanocomposite materials for optical applications \cite{Pinar}. In particular, HCMs can exhibit properties not exhibited by the component mediums from which they arise \cite{Meet}. For example, anisotropic HCMs can rise from isotropic component mediums if the geometries of the inclusions exhibit certain symmetries \cite{MW_JOPA}.

Homogenization formalisms are used to estimate the constitutive parameters
of HCMs \cite{SelPap}. The Maxwell Garnett formalism \cite{MG1904} is perhaps the most widely used of all homogenization formalisms.
Most homogenization formalisms (including the Maxwell Garnett formalism) use \emph{depolarization dyadics} to represent the scattering responses of the inclusions \cite{MAEH}, with the depolarization dyadic being the integrated singularity of the corresponding dyadic Green function \cite{Tai,Faryad}.
 Generally, homogenization formalisms are used for inclusions with simple shapes, e.g., spherical, spheroidal, and ellipsoidal. This is because closed-form expressions
for depolarization dyadics
 are only available
for a limited range of inclusion shapes \cite{M97,Moroz,Osborn,Stoner,MW97,W98,WM02}, with more complex inclusion shapes requiring numerical evaluation \cite{Lee,L98}.

 Recently, the 
range
of inclusion shapes that can be accommodated in homogenization formalisms was extended considerably by the derivation of closed-form expressions for depolarization dyadics for truncated spheres and truncated spheroids. Furthermore, the methodology was extended to truncated ellipsoids \cite{MLarxiv}.
In this paper, these
  recently derived   expressions for depolarization dyadics are used in an implementation of the
   Maxwell Garnett formalism to estimate
  the relative permittivity parameters of HCMs arising from inclusions shaped as truncated spheres and spheroids. The implementation is extended to 
  the case of 
  inclusions with truncated ellipsoidal shapes, with the corresponding depolarization dyadics being evaluated numerically. As regards notational matters: vectors are boldface; dyadics \cite{MAEH,Faryad} are double underlined;   $\epso$ is the free-space permittivity;
$\=I=   \ux\ux+ \uy\uy+ \uz\uz $ is the identity dyadic with  $\ux$, $\uy$, and $\uz$ being the unit vectors aligned with the  coordinate axes of the Cartesian coordinate
system $(x,y,z)$.
 
 \section{Homogenization~--~theoretical matters}

Consider a composite medium 
 comprising  a
 random distribution of identical, similarly oriented, electromagnetically small inclusions surrounded by a host medium.
The inclusion medium has relative permittivity scalar ${\epsi}$, and the host medium  
 ${\epsh}$. 
  The volume fraction of the composite material occupied by the inclusions is denoted by   $\finc$.

 The Maxwell Garnett formalism delivers the following estimate of the relative permittivity dyadic of homogenized composite material:
 \begin{equation} 
 \label{epsMG}
 \=\eps^{MG} = {\epsh}\, \=I +  {\frac{\finc}{\epso}}\,\=a\.\left(\=I-\frac{1}{3\epso\epsh}\finc\,\=a\right)^{-1}.
 \end{equation}
Herein the polarizability density dyadic
 \begin{equation}
\label{pdd-def}
\=a = \epso\left(\epsi-\epsh\right)\left[\=I+\frac{\epsi-\epsh}{\epsh}\=L(\ro)\right]^{-1};
\end{equation}
and the depolarization dyadic
\begin{equation}
\=L(\ro)=    \frac{1}{4 \pi} \iint_{\Si} \hat{\#u}_{\rm n}(\rp) \frac{ \rp-\ro}{\vert\rp-\ro\vert^3} \, d^2 \rp
\end{equation}
is an integral over the
surface $\Si$ of an inclusion, with $\hat{\#u}_{\rm n}$ being the unit outward normal vector to $\Si$
and $\ro$ being the position vector of a point inside the inclusion.  Details leading to Eq.~(\ref{epsMG}) are provided in the
  {\bf Supplementary Material}.

\section{Numerical results and discussion}

We used Eq.~(\ref{epsMG})  to calculate the results presented in this section. Since $\=L(\ro)$ is diagonalizable,
both $\=a$ and $ \=\eps^{MG}$ also are. In addition, all three dyadics have the same set of eigenvectors. This fact
can be used to obtain simple expressions for the three eigenvalues of $ \=\eps^{MG}$, if desired.

\begin{figure}[!htb]
   \begin{subfigure}{.3\textwidth}
\centering
\includegraphics[width=5.5cm]{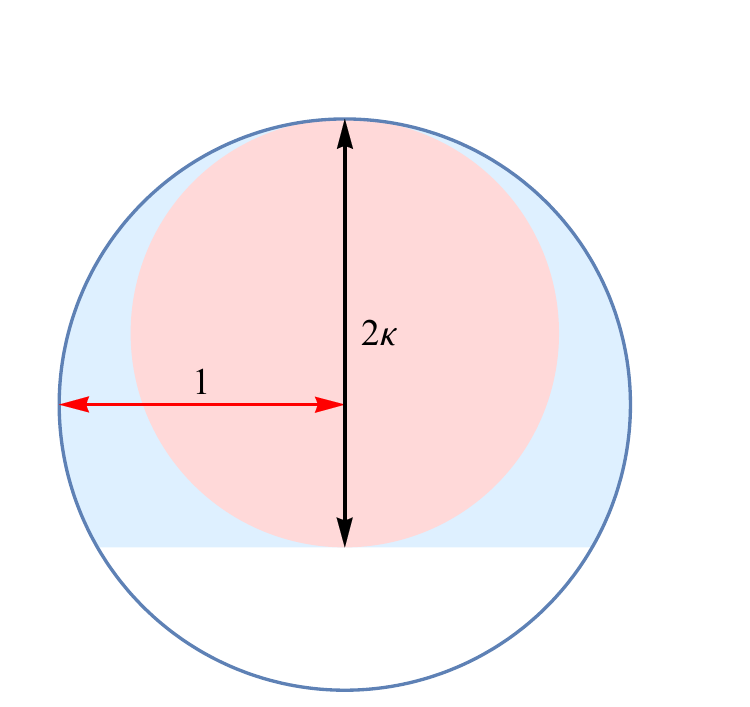} 
  \vspace{0mm}  \hfill
 \caption{\label{Fig1a} Truncated unit sphere with largest inscribed sphere of radius  $\kappa\in(0,1)$.
   }
\end{subfigure} \hspace{1cm}
\begin{subfigure}{.3\textwidth}
\centering
\includegraphics[width=5.5cm]{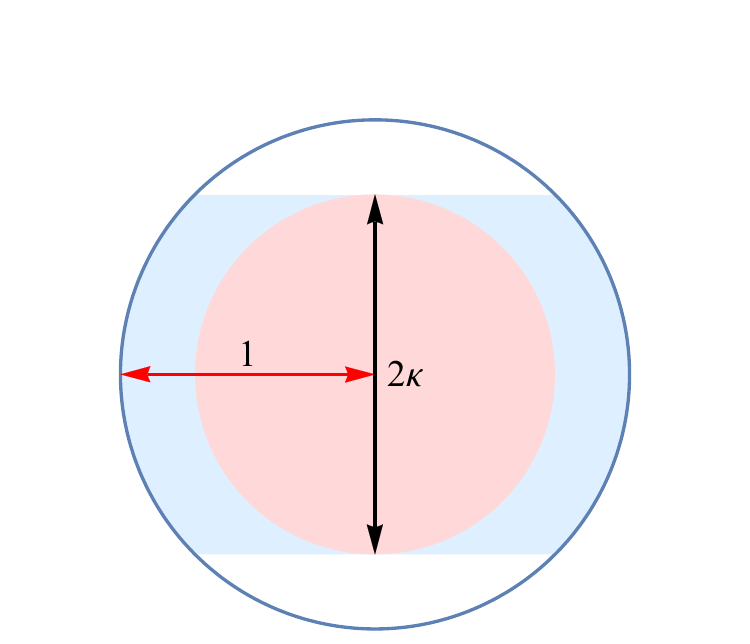} 
  \vspace{0mm}  \hfill
 \caption{\label{Fig1b} Double-truncated unit sphere with largest inscribed sphere of radius $\kappa\in(0,1)$.
   }
\end{subfigure}\\
\begin{subfigure}{.5\textwidth}
\centering
\vspace{0mm}
\includegraphics[height=7.cm]{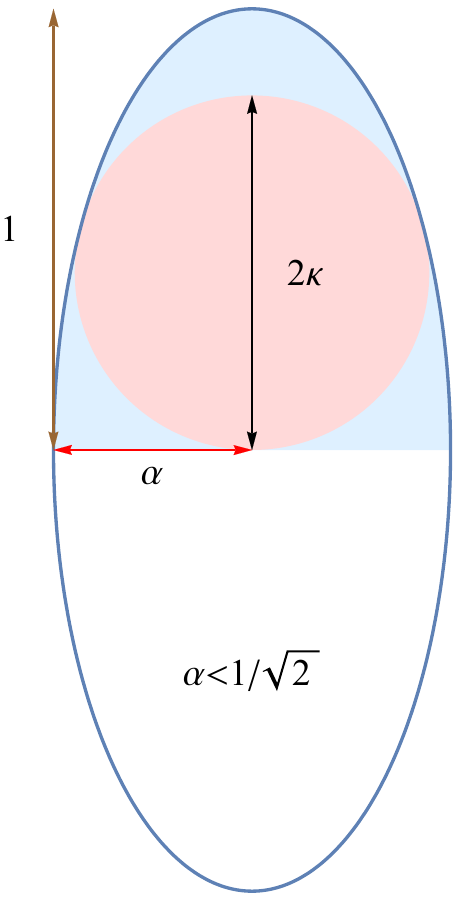} 
\includegraphics[width=5.5cm]{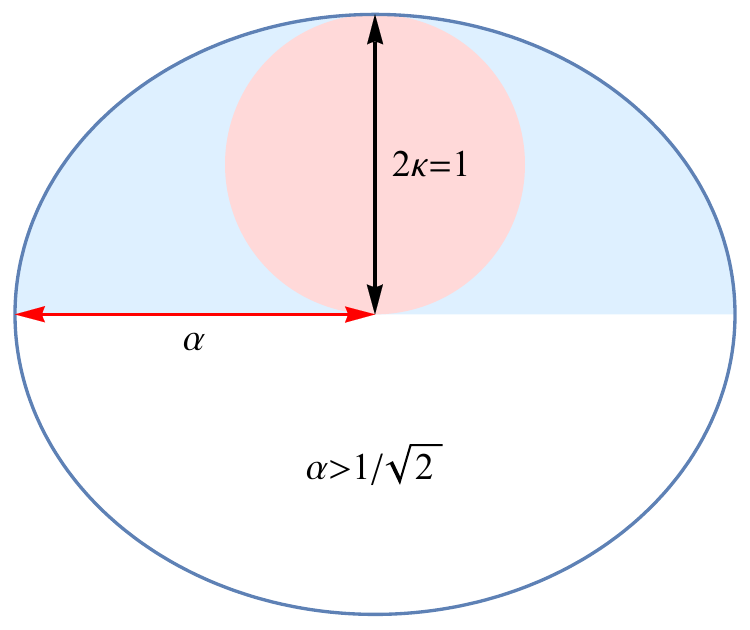} 
  \vspace{0mm}  \hfill
 \caption{\label{Fig1c} Hemispheroid with largest inscribed sphere of radius  $\kappa\in(0,1/2)$ for 
  $\alpha < 1/\sqrt{2}$ (top) and of radius $\kappa = 1/2$ for
 $\alpha > 1/ \sqrt{2}$ (bottom).
   }
\end{subfigure}
\hspace{10mm}
\begin{subfigure}{.5\textwidth}
\centering
\vspace{0mm}
\includegraphics[height=7.cm]{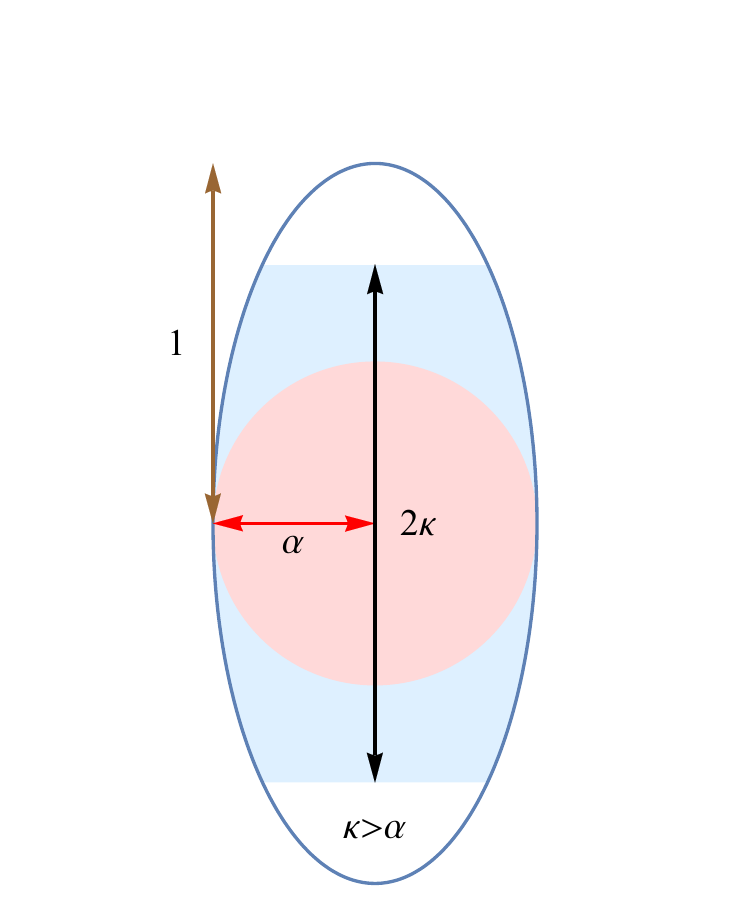} 
\includegraphics[width=7.cm]{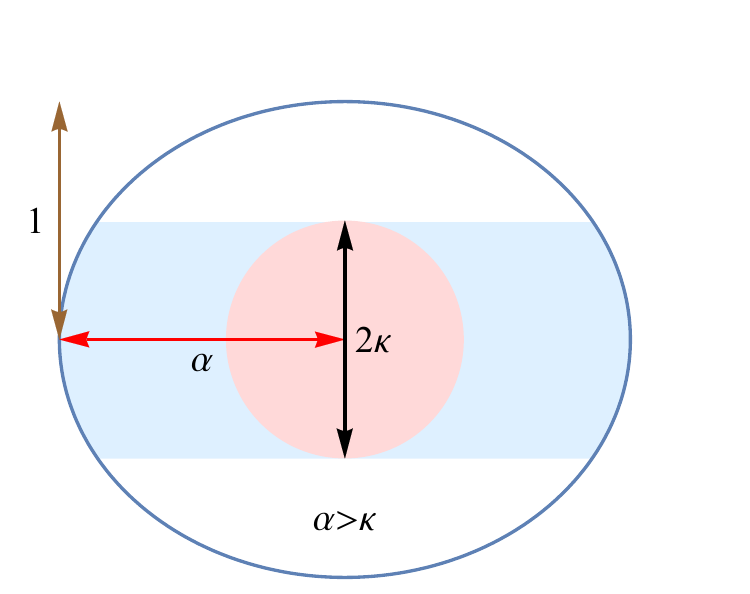} 
  \vspace{0mm}  \hfill
 \caption{\label{Fig1d} Double-truncated spheroid with largest inscribed sphere of radius $\alpha$  for 
  $\alpha < \kappa$ (top) and of radius $\kappa$  for
 $\alpha > \kappa$ (bottom).
   }
\end{subfigure}
 \caption{\label{Fig1} Schematic representations of truncated spheres and spheroids.
   }
\end{figure}

The inclusion shapes were taken to be that of truncated spheres, spheroids, and ellipsoids. For the cases of truncated spheres and spheroids,
closed-form expressions for  $\=L(\ro)$ were recently derived \cite{MLarxiv};
these cases are
 represented schematically in Fig.~\ref{Fig1}. For each type of inclusion shown in this figure,
 $\=L(\ro)$ was derived with
  $\ro$ specifying the  center of the largest sphere inscribed inside the inclusion.

Without loss of generality, the rotational symmetry axis of the truncated spheres, spheroids, and ellipsoids was fixed as the $z$ coordinate axis; i.e., all truncation planes  were parallel to the $xy$  plane.
 The three eigenvectors of $\=L\equiv \=L(\ro)$ are $\ux$, $\uy$, and $\uz$ for all inclusions due to the chosen shapes. Accordingly,
 \begin{equation}
\left.
\begin{array}{l}
\=L  =   {\Lx}\, \ux\ux+ {\Ly}\,\uy\uy+ {\Lz}\,\uz\uz  \vspace{6pt}
\\
 \=\eps^{MG}  =   \epsMGx\, \ux\ux+ \epsMGy\,\uy\uy+ \epsMGz\,\uz\uz 
 \end{array}
 \right\}
 ;
\end{equation}
with ${\Lx}= {\Ly} \equiv \Lt$ and  $\epsMGx= \epsMGy \equiv \epsMGt$ for  truncated spheres and truncated spheroids.

For the purposes of numerical illustration, the representative  values ${\epsh} = 3$, ${\epsi} = 2+ 0.5 i$, and $\finc= 0.3$ were selected,
with higher values of  $\finc$ being  incompatible with the Maxwell Garnett formalism \cite{MAEH,SelPap}. All inclusions are taken to be aligned in parallel, but orientational statistics can be easily incorporated as needed \cite{Lakh}.

 \subsection{Inclusions based on spherical geometry}
The inclusion shape is specified relative to the unit sphere, centered at the coordinate origin.  

\subsubsection{Truncated sphere} \l{Truncated_sphere_sec}

The inclusion shape is that of the upper part of the sphere which is
bounded below by  the plane $z = 1-2 \kappa$, where $0< \kappa <1$, as schematically illustrated in Fig.~\ref{Fig1a}. 
Thus, $\kappa$ represents the radius of the largest sphere that can be inscribed inside the truncated sphere.
In this case, ${\Lz} = 1- 2\Lt$ and
\begin{equation}
\Lt =  \frac{ \kappa  }{6 \le 1 - \kappa\ri^3 \tau}\,\lec 6 - \tau \le 3  - 3 \kappa  + \kappa^2 \ri - \kappa \les 11 - 3 \kappa \le 3- \kappa 
\ri\ris \ric\,,
\end{equation}
where
\begin{equation}
\tau = \sqrt{\le 4 - 3 \kappa \ri \kappa}\,.
\end{equation}

\begin{figure}[!htb]
\centering
\includegraphics[width=8cm]{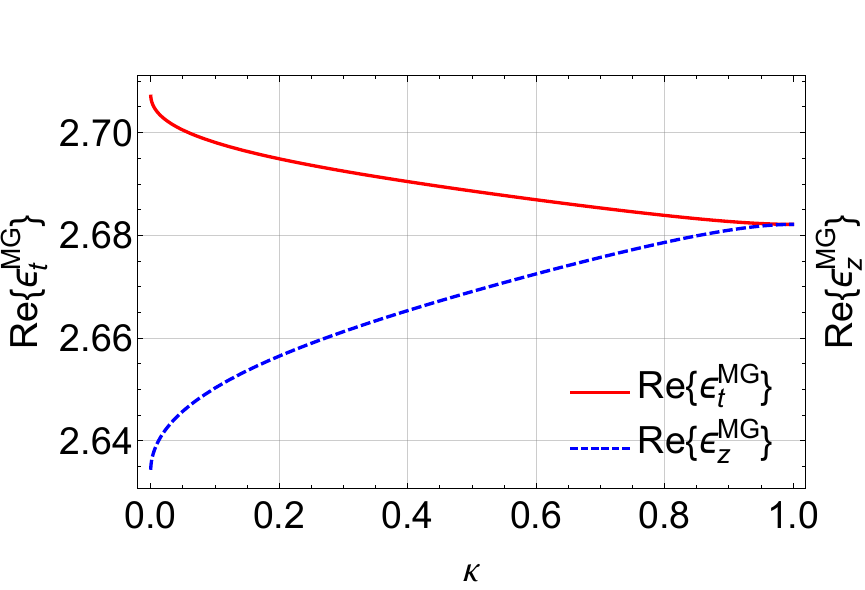} 
\includegraphics[width=8cm]{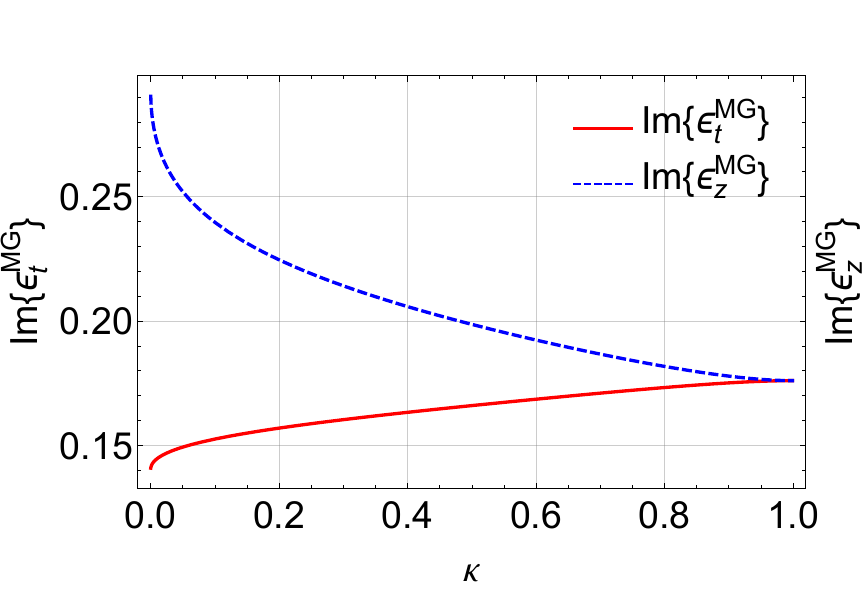} 
  \vspace{0mm}  \hfill
 \caption{\label{Fig2} Real and imaginary parts of $\epsMGt$ and $\epsMGz$ plotted against  $\kappa\in(0,1)$ for truncated
 spherical inclusions. 
   }
\end{figure}

Plots of the
real and imaginary parts of the 
   relative permittivity parameters $\epsMGt$ and $\epsMGz$ of the HCM versus $\kappa$ are provided in Fig.~\ref{Fig2}. As $\kappa$ increases,
the values of $\mbox{Re}\lec \epsMGt \ric$ and 
$\mbox{Im}\lec \epsMGz \ric$
decrease monotonically  whereas the values of 
$\mbox{Re}\lec \epsMGz \ric$ and 
$\mbox{Im}\lec \epsMGt \ric$
 increase monotonically. For all $\kappa < 1$, $\mbox{Re}\lec \epsMGt \ric> 
 \mbox{Re}\lec \epsMGz \ric$ and $\mbox{Im}\lec \epsMGz \ric> 
 \mbox{Im}\lec \epsMGt \ric$.
The differences between $\mbox{Re}\lec \epsMGt \ric$ and
 $\mbox{Re}\lec \epsMGz \ric$ and between 
$\mbox{Im}\lec \epsMGt \ric$ and
 $\mbox{Im}\lec \epsMGz \ric$
decrease monotonically as $\kappa$ increases, and these differences vanish in the limit $\kappa \to 1$. That is, the degree of anisotropy that the HCM exhibits decreases monotonically as $\kappa$ increases and  the HCM becomes isotropic
 in the limit $\kappa \to 1$.

\subsubsection{Double-truncated sphere}

The inclusion shape is that of the middle part of the sphere which is
bounded below by  the plane $z = - \kappa$ and bounded above
by the plane $z = \kappa$,  where $0< \kappa <1$, as schematically illustrated in Fig.~\ref{Fig1b}. 
Thus, $\kappa$  represents the radius of the largest sphere that can be inscribed inside the double-truncated sphere.
In this case,  ${\Lz}=1-2\Lt$ with
\begin{equation}
\Lt= \frac{ \le 3- \kappa^2 \ri \kappa}{6}\,.
\end{equation}
 
\begin{figure}[!htb]
\centering
\includegraphics[width=8cm]{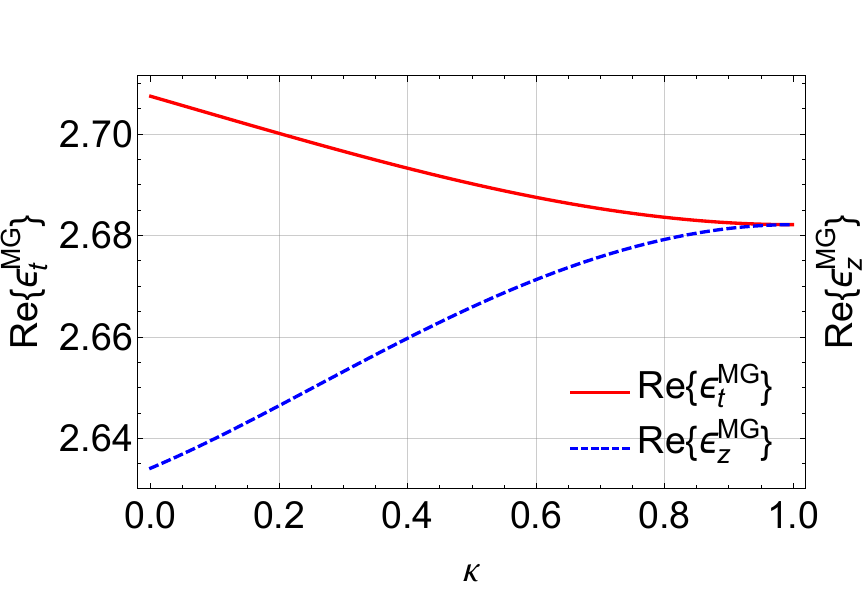} 
\includegraphics[width=8cm]{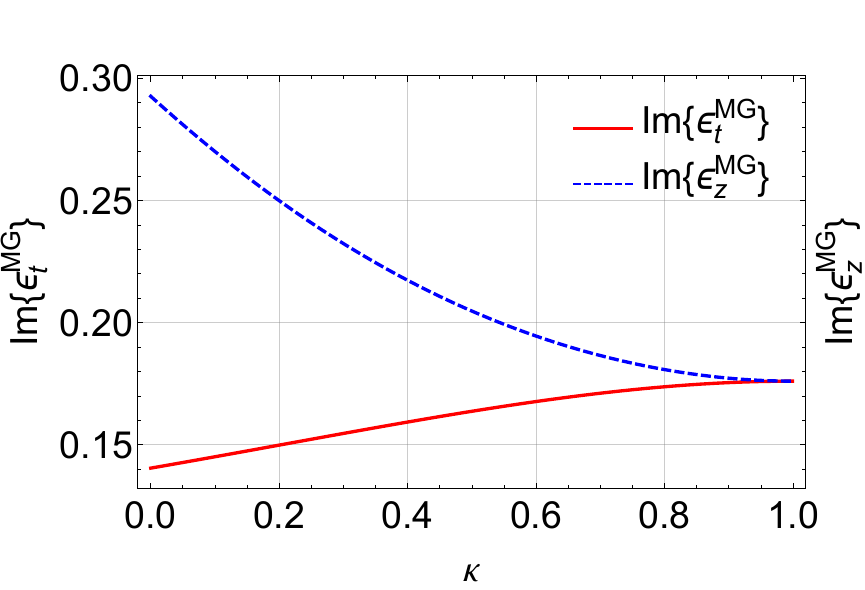} 
  \vspace{0mm}  \hfill
 \caption{\label{Fig3} Real and imaginary parts of
 $\epsMGt$ and $\epsMGz$ plotted  against  $\kappa\in(0,1)$ for double-truncated spherical
 inclusions.  
   }
\end{figure}

In Fig.~\ref{Fig3}, 
the
real and imaginary parts of the 
   relative permittivity parameters $\epsMGt$ and $\epsMGz$ of the HCM
 are plotted against $\kappa$.
As is the case for the truncated sphere presented in \S\ref{Truncated_sphere_sec}, for the double-truncated sphere
the HCM is isotropic in the limit $\kappa \to 1$ and exhibits an increasing degree of anisotropy as $\kappa$ decreases.
The plots of 
the real and imaginary parts of  $\epsMGt$ and $\epsMGz$
 for the double-truncated sphere
in Fig.~\ref{Fig3} are similar to the corresponding plots in Fig.~\ref{Fig2} for the truncated sphere. Differences are
most conspicuous in the regime of small values of $\kappa$ 
wherein 
the real and imaginary parts of  $\epsMGt$ and $\epsMGz$
change less rapidly  as $\kappa$ increases
for the double-truncated sphere.

\subsection{Inclusions based on spheroidal geometry}

The inclusion shape is specified relative to the  spheroid
\begin{equation}
\frac{x^2 + y^2}{\alpha^2}  +  z^2 \leq 1,
\end{equation}
with equatorial radius  $\alpha>0$.

\subsubsection{Hemispheroid }

The inclusion shape is that of the upper half of the spheroid; this shape  lies between 
  the planes $z = 0$ and $z=1$, as schematically illustrated in Fig.~\ref{Fig1c}. 
The radius of the largest sphere that can be inscribed inside the hemispheroid
is $\kappa$; we have $\kappa = \alpha \sqrt{1-\alpha^2} \in (0, 1/2)$ for $\alpha < 1/ \sqrt{2}$ while $\kappa = 1/2$ for 
$\alpha > 1 / \sqrt{2}$.
For $\alpha < 1 / \sqrt{2}$, we have
\begin{eqnarray} 
\Lt &=& \frac{1}{4 \nu \gamma^4} \left\{
1+ \nu + \alpha^2 \les \alpha \le \alpha + \sqrt{2- \alpha^2} \ri- \nu - 3 \ris
\right.
\nonumber \\
&& \left.
- \le\alpha \gamma \ri^2 \sqrt{2 - \alpha^2} 
\log \displaystyle{\frac{\le 1- \alpha^2 + \gamma \ri
\le 1- \alpha \ri}{\alpha \le \nu - \gamma \ri}}\right\},
\end{eqnarray}
where
\begin{equation}
\left.
\begin{array}{l}
 \nu =  \sqrt{2 - 3 \alpha^2 + \alpha^4}
 \vspace{6pt}\\
 \gamma = \sqrt{1-\alpha^2 }
\end{array}
\right\}
;
\end{equation}
 for $ \alpha > 1 / \sqrt{2}$, we have
\begin{eqnarray} 
\Lt &=& \displaystyle{\frac{3 \le 1+ \sqrt{1+4\alpha^2} \ri
+ \alpha^2 \le 10 - 8 \sqrt{1+4 \alpha^2} \ri - 8 \alpha^4}{4 \gamma^2  
\le    3 +  8 \alpha^2 - 16 \alpha^4 \ri }
}
\nonumber \\ &&
\displaystyle{
+ \frac{\alpha^2 \log
\displaystyle{ \frac{1-2\alpha^2 + \gamma}
{\sqrt{1+ 3 \alpha^2 - 4 \alpha^4 }-1}}}{4 \gamma^3 }
}\,.
\end{eqnarray}
For all $\alpha$, ${\Lz} = 1 - 2\Lt$.

\begin{figure}[!htb]
\centering
\includegraphics[width=8cm]{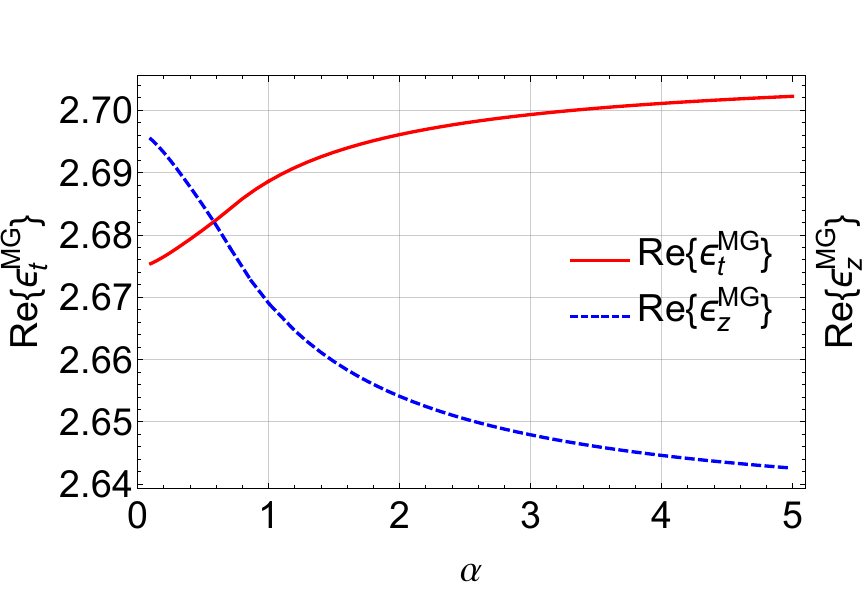} 
\includegraphics[width=8cm]{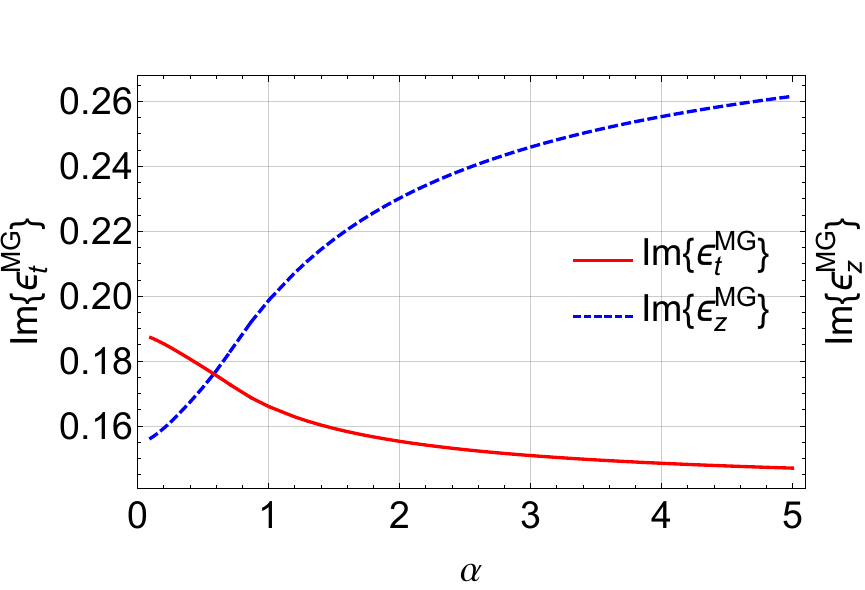} 
  \vspace{0mm}  \hfill
 \caption{\label{Fig4} Real and imaginary parts of
 $\epsMGt$ and $\epsMGz$ plotted against $\alpha \in (0,5)$ for hemispheroidal
 inclusions.  
   }
\end{figure}

Plots of 
the
real and imaginary parts of the 
  relative permittivity parameters $\epsMGt$ and $\epsMGz$ of the HCM
 versus $\alpha$ are displayed in Fig.~\ref{Fig4}.
As $\alpha$ increases,
the values of 
$\mbox{Re}\lec \epsMGz \ric$
and $\mbox{Im}\lec \epsMGt \ric$
 decrease monotonically,  whereas the values of 
 $\mbox{Re}\lec \epsMGt\ric$
 and
 $\mbox{Im}\lec \epsMGz \ric$
  increase monotonically. The value $\alpha = 0.569$
  is noteworthy because for this value $ \epsMGt =\epsMGz $
  and the HCM is consequently isotropic. As $\alpha$ 
deviates farther from 0.569,
  the degree of anisotropy exhibited by the HCM increases.
Results for   hemispherical inclusions are obtained for $\alpha=1$.

\subsubsection{Double-truncated spheroid}

The inclusion shape is that of the middle part of the spheroid which is
bounded below by  the plane $z = - \kappa$ and bounded above
by the plane $z = \kappa$, 
with $0 < \kappa < 1$,
as schematically illustrated in Fig.~\ref{Fig1d}. 
The radius of the largest sphere that can be inscribed inside the double-truncated spheroid
is $\alpha$ for $\kappa > \alpha$
and the radius is $\kappa$ for $\kappa < \alpha$.
In this case
\begin{equation}
\Lt =\frac{- \nu + \alpha^2 \tan^{-1} \nu}{2 \le\alpha^2 -1 \ri^{3/2}}
\end{equation}
with
\begin{equation}
\nu = \kappa \sqrt{\frac{\alpha^2 -1}{\alpha^2- \kappa^2 \le \alpha^2-1 \ri}};
\end{equation}
and again ${\Lz} = 1- 2 \Lt$. 

\begin{figure}[!htb]
\centering
\includegraphics[width=7cm]{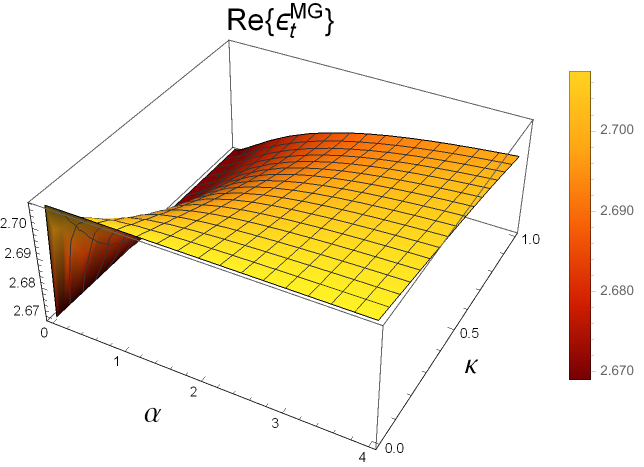}
\includegraphics[width=7cm]{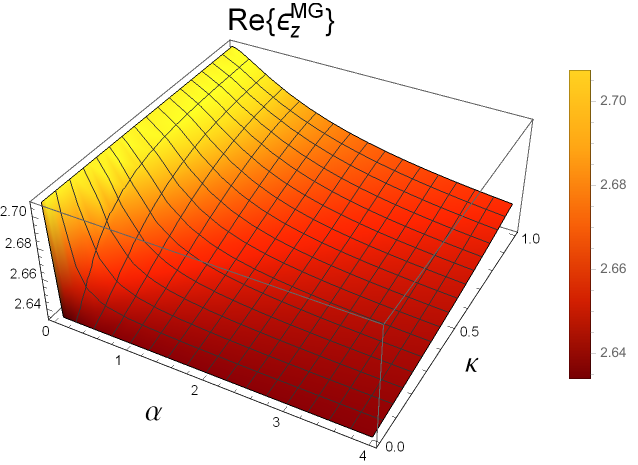} \\
\includegraphics[width=7cm]{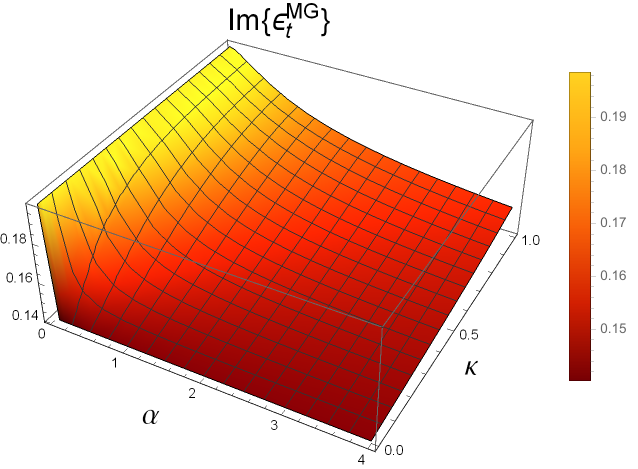}
\includegraphics[width=7cm]{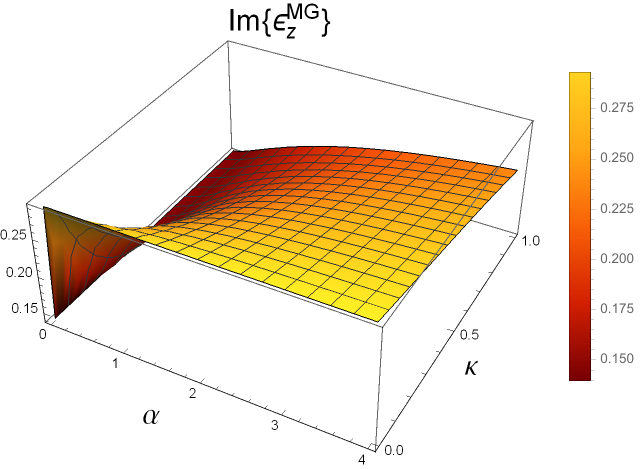}
  \vspace{0mm}  \hfill
 \caption{\label{Fig5}  Real and imaginary parts of
 $\epsMGt$ and $\epsMGz$   plotted against $\alpha\in(0,4]$ and  $\kappa\in(0,1)$ for     double-truncated spheroidal inclusions.
   }
\end{figure}

In Fig.~\ref{Fig5}, the 
real and imaginary parts of the 
   relative permittivity parameters $\epsMGt$ and $\epsMGz$ of the HCM are plotted against $\alpha$ and $\kappa$.
As $\alpha$ increases,  $\mbox{Re}\lec \epsMGt \ric$
and $\mbox{Im}\lec \epsMGz \ric$
 increase monotonically, and 
 $\mbox{Re}\lec \epsMGz \ric$
and $\mbox{Im}\lec \epsMGt \ric$
 decrease monotonically, 
for all values of $\kappa$. As $\kappa$ increases, 
$\mbox{Re}\lec \epsMGt \ric$
and $\mbox{Im}\lec \epsMGz \ric$
 decrease monotonically, and
 $\mbox{Re}\lec \epsMGz \ric$
and $\mbox{Im}\lec \epsMGt \ric$
 increase monotonically,
 for all values of $\alpha$.
In the limit $\kappa \to 0$, the values of
$ \epsMGt $
and $\epsMGz $ are largely independent of $\alpha$, except at the smallest 
values of $\alpha$.

\subsection{Inclusions based on ellipsoidal geometry}

The inclusion shape is specified relative to the ellipsoid
\begin{equation}
\frac{x^2}{\alpha^2} + \frac{y^2}{\beta^2} + z^2 \leq 1,
\end{equation}
with semi-axis lengths $\alpha>0$ and $\beta>0$.

\subsubsection{Hemi-ellipsoid }

The inclusion shape is that of the upper half of the ellipsoid; this shape   lies between 
  the planes $z = 0$ and $z=1$. 
  The 
 largest  sphere inscribed inside the hemi-ellipsoid
has radius
\begin{equation}
\kappa = \left\{  \begin{array}{lcr}
1/2,  & \mbox{for}& \alpha > 1 / \sqrt{2}, \quad \beta > 1 / \sqrt{2} \vspace{8pt}\\
\alpha \sqrt{1- \alpha^2}& \mbox{for} & \alpha \leq 1 / \sqrt{2}, \quad \alpha < \beta
 \vspace{8pt}\\
\beta \sqrt{1- \beta^2}& \mbox{for} & \beta \leq 1 / \sqrt{2}, \quad \beta < \alpha
\end{array}\right.
.
\end{equation}
The dyadic $\=L$ has to be determined by numerical integration.

\begin{figure}[!htb]
\centering
\includegraphics[width=7cm]{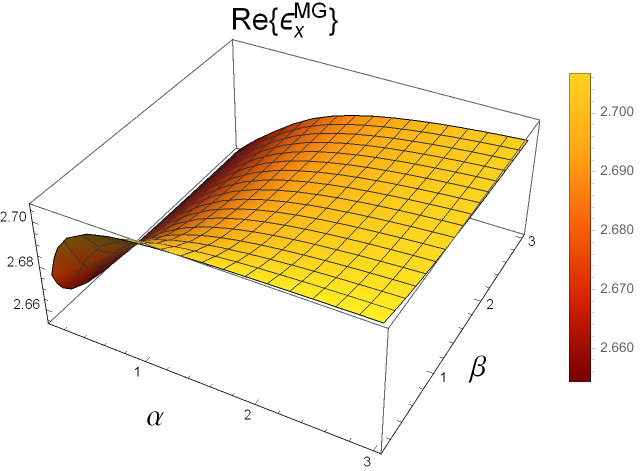} 
\includegraphics[width=7cm]{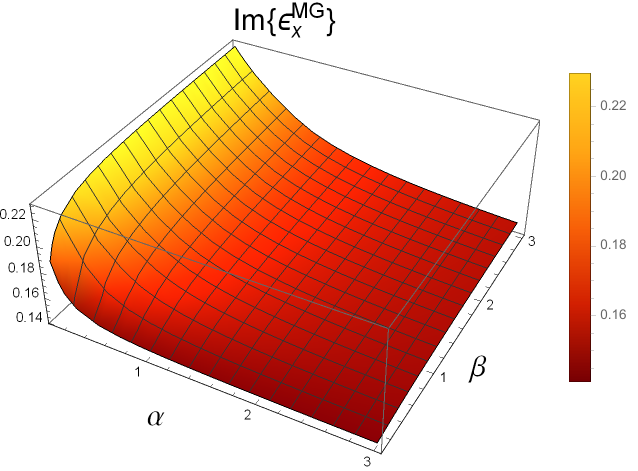} \\
\includegraphics[width=7cm]{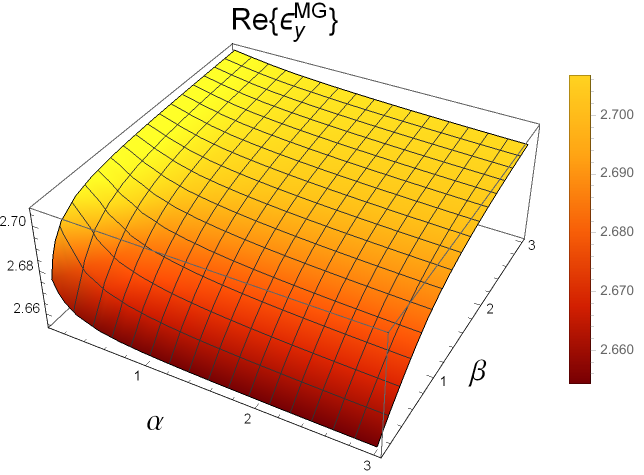} 
\includegraphics[width=7cm]{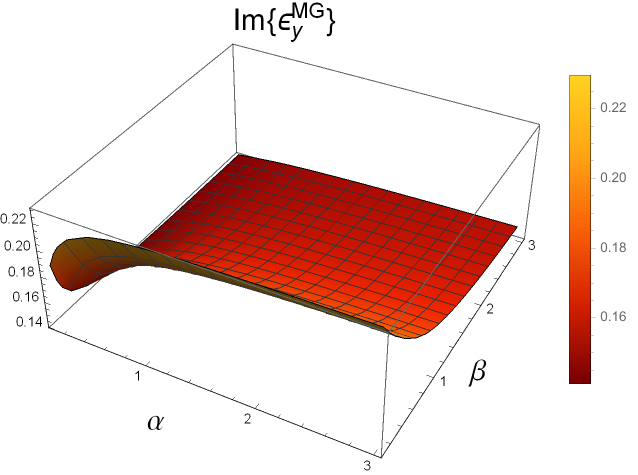} \\
\includegraphics[width=7cm]{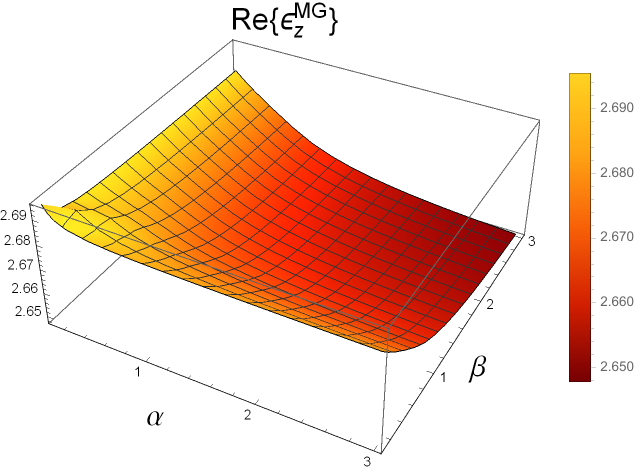} 
\includegraphics[width=7cm]{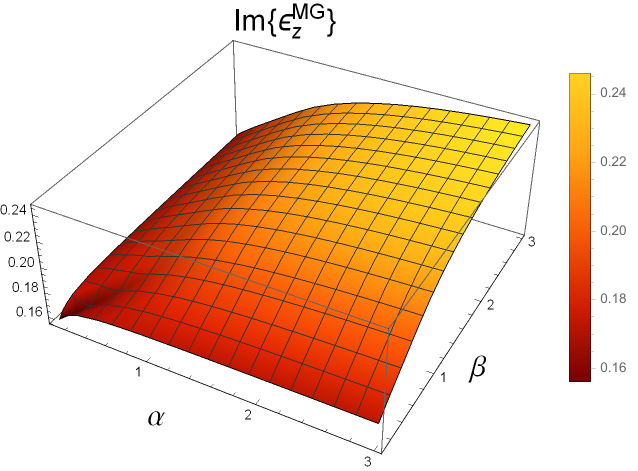} 
 \caption{\label{Fig6}
  Real and imaginary parts of
 $\epsMGx$, $\epsMGy$, and $\epsMGz$
   plotted against $\alpha\in(0,3]$ and $\beta\in(0,3]$
   for   hemi-ellipsoidal inclusions.
   }
\end{figure}

In Fig.~\ref{Fig6},
plots of the 
real and imaginary parts of the 
   relative permittivity parameters $\epsMGx$,
 $\epsMGy$,
  and $\epsMGz$ of  the HCM
 versus $\alpha$ and $\beta $ are displayed.
 The real and imaginary parts of
each of  $\epsMGx$,
 $\epsMGy$,
  and $\epsMGz$ vary smoothly as $\alpha$ and $\beta$ increase. In particular,
$\mbox{Re} \lec \epsMGx \ric$
 increases markedly 
 and
$\mbox{Im} \lec \epsMGx \ric$
 decreases markedly 
 as $\alpha$ increases but both are relatively insensitive to changes in $\beta$; 
$\mbox{Re} \lec \epsMGy \ric$
 increases markedly 
 and
$\mbox{Im} \lec \epsMGy \ric$
 decreases markedly 
 as $\beta$ increases but both are relatively insensitive to changes in $\alpha$; 
 and
 $\mbox{Re} \lec \epsMGz \ric$
 decreases markedly as both  $\alpha$  and $\beta$ increase whereas
$\mbox{Im} \lec \epsMGz \ric$
 increases markedly 
 as both  $\alpha$  and $\beta$ increase.

\subsubsection{Double-truncated ellipsoid}

The inclusion shape is that of the middle part of the ellipsoid which is
bounded below by  the plane $z = - \kappa$ and bounded above
by the plane $z = \kappa$, 
with $0 < \kappa < 1$. Again, $\=L$ has to be determined by numerical integration.

\begin{figure}[!htb]
\centering

\includegraphics[width=7cm]{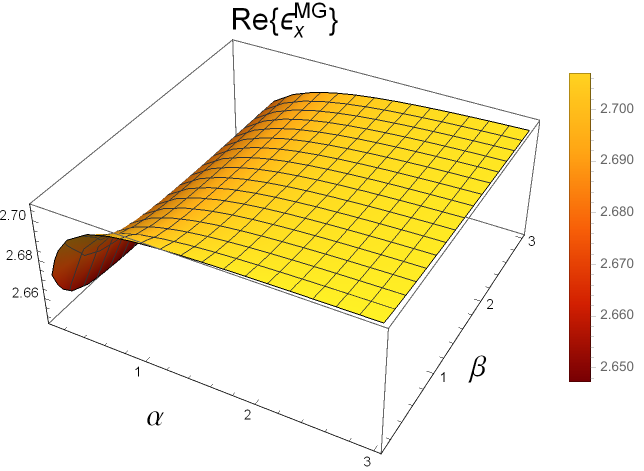}  \hspace{5mm}
 \includegraphics[width=7cm]{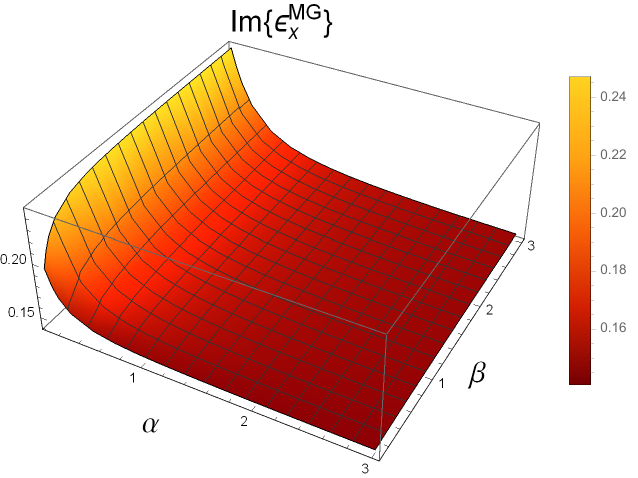}\\
\includegraphics[width=7cm]{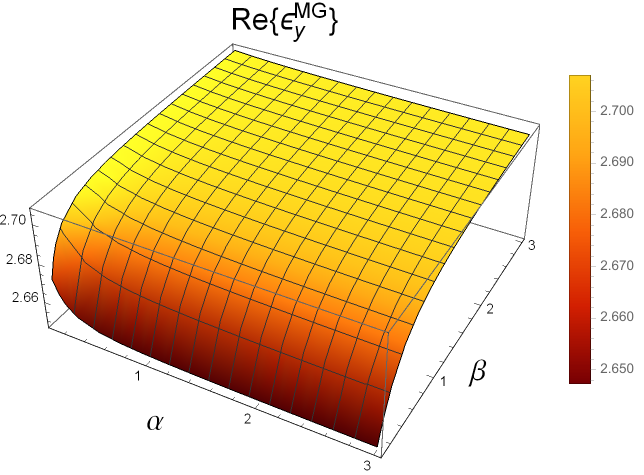} \hspace{5mm}
\includegraphics[width=7cm]{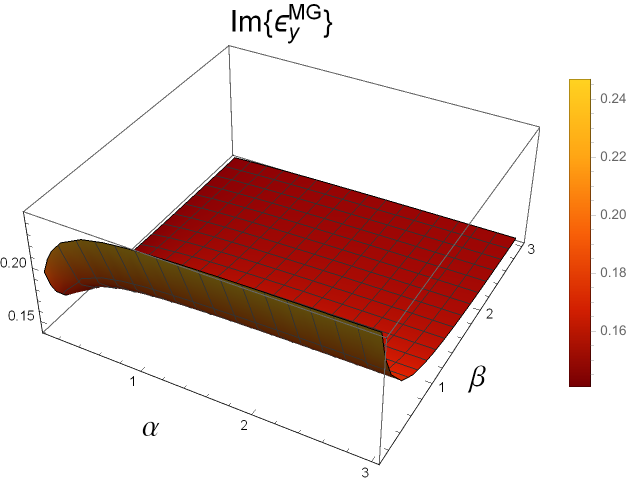} \\
\includegraphics[width=7cm]{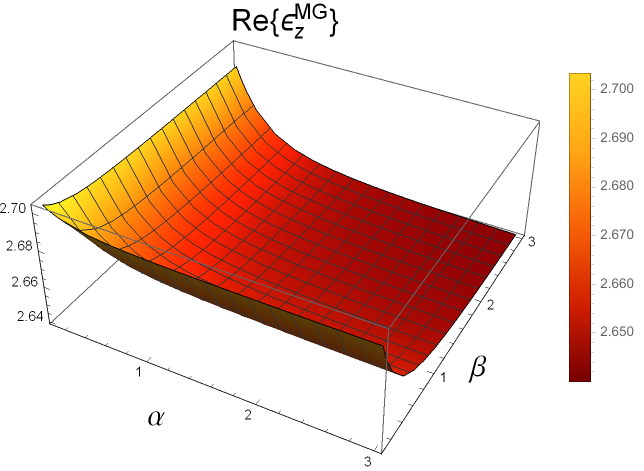}  \hspace{5mm}
 \includegraphics[width=7cm]{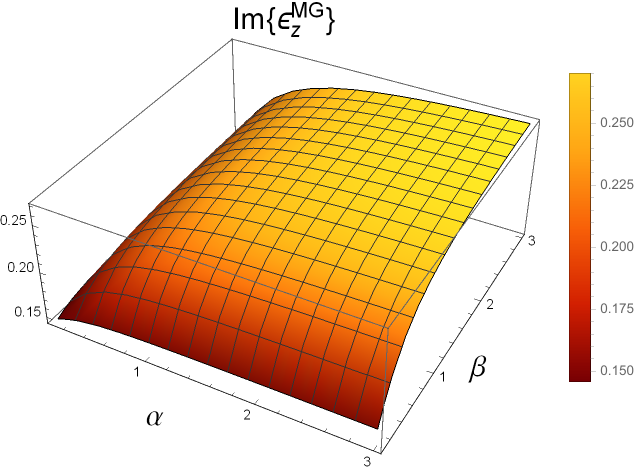}
 \caption{\label{Fig7}
 Real and imaginary parts of
 $\epsMGx$, $\epsMGy$, and $\epsMGz$ plotted against  $\alpha\in(0,3]$ and $\beta\in(0,3]$  for 
 double-truncated ellipsoidal inclusions   with $\kappa = 0.3$.}
\end{figure}

For $\kappa =0.3 $, 
plots of 
the 
real and imaginary parts of the 
   relative permittivity parameters $\epsMGx$,
 $\epsMGy$,
  and $\epsMGz$ of the HCM versus $\alpha$ and $\beta $ are provided in Fig.~\ref{Fig7}.
The plots in Fig.~\ref{Fig7} for the double-truncated ellipsoid 
are qualitatively similar to those in Fig.~\ref{Fig6} for the hemi-ellipsoid.
The effects of varying $\kappa$ are most appreciable at low values of $\alpha$ for 
both  $\mbox{Re} \lec \epsMGx \ric$ and  $\mbox{Im} \lec \epsMGx \ric$, at low values
of $\beta$ for
both  $\mbox{Re} \lec \epsMGy \ric$ and  $\mbox{Im} \lec \epsMGy \ric$, and at low values of both $\alpha$ and $\beta$ for both  $\mbox{Re} \lec \epsMGz \ric$ and  $\mbox{Im} \lec \epsMGz \ric$. From further numerical studies (not presented graphically here), the values of
$\epsMGx$,
 $\epsMGy$,
  and $\epsMGz$   are 
generally much  more sensitive to variations in $\alpha$ and $\beta$ than they are to variations in $\kappa \in (0,1)$.

 \section{Closing remarks}
 
 In this paper, the Maxwell Garnett formalism was implemented to estimate the relative permittivity parameters of HCMs arising from
 an isotropic host medium impregnated with isotropic
  inclusions of relatively complex shapes. The inclusion shapes considered
 were truncated spheres, spheroids, and ellipsoids. In so doing, the anisotropy of the HCM was related to the geometry of the inclusions: in general, the more  the shape of the inclusions deviates from spherical, the greater is the degree of anisotropy  exhibited by the HCM.
 The availability of closed-form expressions for depolarization dyadics for inclusions shaped as  truncated spheres and spheroids \cite{MLarxiv}  increases considerably the reach of the Maxwell Garnett formalism, allowing more complex HCMs to be theoretically analyzed. Additionally, this formalism can be used for more realistic modeling of poultry dust \cite{Poultry-dust}, house dust \cite{House-dust1,House-dust2}, cements \cite{Cement},
 volcanic ash \cite{Volcanic-ash1,Volcanic-ash2}, mineral dust \cite{Mineral-dust,Martian-dust}, and interstellar dust \cite{Interstellar-dust}, 
 which are presently modeled very simply as agglomerations of complete ellipsoids \cite{Spheroidal-shape1,Spheroidal-shape2}. Although we considered all inclusions in a composite medium to be identical and aligned in parallel, both of those restrictions can be easily surmounted
 as shown elsewhere \cite{Lakh,Perc}.
 
 The Maxwell Garnett formalism is usually implemented for cases involving spherical, spheroidal, and ellipsoidal inclusions. In such cases, the derivations of the corresponding polarizability density dyadics  rest upon the assumption that electromagnetic  field within each inclusion is spatially uniform. This assumption is physically reasonable provided that the inclusions are electrically small. In the preceding sections, this assumption was extended to truncated spheres, truncated spheroids, and truncated ellipsoids. For such truncated inclusions, this assumption is also physically reasonable provided that the inclusions are sufficiently small and that their shapes do not deviate too much from  non-truncated spheres, spheroids, and ellipsoids.

Finally, all homogenization formalisms~--~including the Maxwell Garnett formalism~--~involve various assumptions and approximations. The acid test of any homogenization formalism can only come from comparison with suitable experimental data.

\section*{Acknowledgments}
TGM was supported  by
EPSRC (grant number EP/V046322/1).  AL   was supported by     the Charles Godfrey Binder Endowment at Penn State.

\section*{Declaration of competing interest}
The authors declare that they have no known competing financial interests or personal relationships that could have
appeared to influence the work reported in this paper.

\section*{Data availability}
All data  analyzed in this paper are included in this paper.

\newpage

\newpage

\vspace{10mm}

\vspace{10mm}

\vspace{10mm}

\newpage

\vspace{10mm}

\newpage

\newpage

\begin{center}

\LARGE{ {\bf \textsc{Supplementary Material:}\\ Anisotropic homogenized composite mediums arising from  truncated spheres, spheroids, and ellipsoids
}}
\end{center}
\begin{center}
\vspace{10mm} \large
 
 {Tom G. Mackay}\footnote{E--mail: T.Mackay@ed.ac.uk.}\\
{\em School of Mathematics and
   Maxwell Institute for Mathematical Sciences\\
University of Edinburgh, Edinburgh EH9 3FD, UK}\\
and\\
 {\em NanoMM~---~Nanoengineered Metamaterials Group\\ Department of Engineering Science and Mechanics\\
The Pennsylvania State University, University Park, PA 16802--6812,
USA}
 \vspace{3mm}\\
 {Akhlesh  Lakhtakia}\\
 {\em NanoMM~---~Nanoengineered Metamaterials Group\\ Department of Engineering Science and Mechanics\\
The Pennsylvania State University, University Park, PA 16802--6812, USA}

\normalsize

\end{center}

\setcounter{section}{0}
\setcounter{equation}{0}

\section{Introduction}

In this document, the derivations of the polarizability density dyadic and the
 Maxwell Garnett formalism are outlined, no novelty being claimed for their derivations.
 The electrodynamic provenance of these theoretical constructions is emphasized.

 As regards notational matters: vectors are boldface; dyadics \cite{xMAEH,xFaryad} are double underlined;  $i = \sqrt{-1}$;  the   angular frequency is $\omega$  and an $\exp(-i\omega t)$
   dependence on time $t$ is implicit; $\epso$ is the free-space permittivity,
   $\muo$ is the free-space  permeability, and $\ko=\omega\sqrt{\epso\muo}$ is the free-space wavenumber;
$\=I=   \ux\ux+ \uy\uy+ \uz\uz $ is the identity dyadic;  $\#r=x\ux+y\uy+z\uz$
is the position vector; and $\ux$, $\uy$, and $\uz$ are the unit vectors aligned with the  coordinate axes of the Cartesian coordinate
system $(x,y,z)$.
 
\section{Polarizability density dyadic} \l{PDD_sec}

A crucial ingredient in homogenization formalisms for particulate composite mediums is the polarizability density dyadic
of a solitary inclusion immersed in a host medium. Sometime in 1848,  Mossotti \cite{xMossotti}
conceptualized the existence of the polarizability
density scalar of an electrically small, isotropic-dielectric sphere in free space. In this section, we present a clear derivation
of the polarizability density dyadic of an electrically small, isotropic-dielectric inclusion of quite general shape embedded in an isotropic
dielectric host medium. No novelty is implied for this derivation.

Let the closed surface $\Si$ separate the  electrically small region $\Vi$  from the region $\Vh$ that extends to infinity in all directions. The shape
of $\Vi$ must be such that
the unit outward normal
$\hat{\#u}_{\rm n}(\#r)$  can be unambiguously identified at every point $\#r \in \Si$;
if necessary, wedges and vertexes on $\Si$ can be rounded off slightly to overcome this restriction.

Whereas $\Vi$ is occupied by a homogeneous dielectric medium of relative permittivity $\epsi(\omega)$, $\Vh$ is occupied by a homogeneous
dielectric medium of relative permittivity $\epsh(\omega)=\nh^2(\omega)$. 

Suppose that the region $\Vi$ is irradiated by a monochromatic electromagnetic field phasor $\#E_{\rm s}(\#r,\omega)$ whose source is confined
to the bounded region $\Vs\subset\Vh$  far from $\Vi$. Then,
 the actual electric field phasor
at any location $\#r\notin\Vs$ is given exactly by \cite{xJRNIST,xChew,xSancer}
\begin{equation}
\label{xreq1}
\#E(\#r,\omega)= \#E_{\rm s}(\#r,\omega)+\ko^2\left[\epsi(\omega)-\epsh(\omega)\right]\iiint_{\Vi}\left[\=G\host(\#r,\rp;\omega)\.\#E(\rp,\omega)\right]\,d^3\rp\,,\quad \#r\notin\Vs\,,
\end{equation}
where
\begin{equation}
\label{xreq2}
\=G\host(\#r,\rp;\omega)=\left[\=I + \frac{\nabla\nabla}{\ko^2\epsh(\omega)}\right]\frac{\exp\left[i\ko\nh(\omega)\vert\#r-\rp\vert\right]}{4\pi\vert\#r-\rp\vert}\,.
\end{equation}
 Setting $\#r=\ro$ in Eq.~(\ref{xreq1}), where
$\ro\in\Vi$  but $\ro\notin\Si$, we get
\begin{equation}
\label{xreq1p}
\#E(\ro,\omega)= \#E_{\rm s}(\ro,\omega)+\ko^2\left[\epsi(\omega)-\epsh(\omega)\right]\iiint_{\Vi}\left[\=G\host(\ro,\rp;\omega)\.\#E(\rp,\omega)\right]\,d^3\rp\,,\quad \ro\in\Vi\,.
\end{equation}

The integral on the right side of  Eq.~(\ref{xreq1p}) is improper due to the appearance of the term $\vert\ro-\rp\vert^{-3}$ in the integrand
\cite{xKellogg}. In order to overcome this difficulty in a ``unified and consistent" manner \cite{xWang}, the term $\=G_{\rm P}(\ro,\rp;\omega)\.\#E(\ro,\omega)$ is added to and subtracted
from that integrand to obtain  \cite{xFikioris}
\begin{eqnarray}
\nonumber
&&
\#E(\ro,\omega)= \#E_{\rm s}(\ro,\omega)+\ko^2\left[\epsi(\omega)-\epsh(\omega)\right]
\iiint_{\Vi}\left[\=G\host(\ro,\rp;\omega)\.\#E(\#\rp,\omega)-\=G_{\rm P}(\ro,\rp;\omega)\.\#E(\ro,\omega)
\right]\,d^3\rp
\\[5pt]
&&\qquad
+\ko^2\left[\epsi(\omega)-\epsh(\omega)\right]
\left[\iiint_{\Vi} \=G_{\rm P}(\ro,\rp;\omega)\,d^3\rp\right]\.\#E(\ro,\omega)
\,,\quad \ro\in\Vi\,,
\label{xreq3}
\end{eqnarray}
where 
\begin{equation}
\label{xreq4}
\=G_{\rm P}(\#r,\rp;\omega)= \frac{1}{\ko^2\epsh(\omega)} \nabla\nabla\frac{1}{4\pi\vert\#r-\rp\vert}\,.
\end{equation}

In order to treat the second integral on the right side of Eq.~(\ref{xreq3}), first the identity $\nabla\nabla\left(\vert\#r-\rp\vert^{-1}\right)=-\nabla\nabla^\prime\left(\vert\#r-\rp\vert^{-1}\right)$ is invoked and then the Gauss theorem is used to obtain \cite{xWang,xJRNIST}
\begin{equation}
\#E(\ro,\omega)= \#E_{\rm s}(\ro,\omega)+\ko^2\left[\epsi(\omega)-\epsh(\omega)\right]\left[\#M(\ro,\omega)-\frac{1}{\ko^2\epsh(\omega)}\=L (\ro)
\.\#E(\ro,\omega)\right]
\,,\quad \ro\in\Vi\,,
\label{xreq5}
\end{equation}
where 
\begin{equation}
\#M (\ro,\omega)= \iiint_{\Vi}\left[\=G\host(\ro,\rp;\omega)\.\#E(\#\rp,\omega)-\=G_{\rm P}(\ro,\rp;\omega)\.\#E(\ro,\omega)
\right]\,d^3\rp\,,
\end{equation}
and
\begin{equation}
\=L(\ro)=    \frac{1}{4 \pi} \iint_{\Si} \hat{\#u}_{\rm n}(\rp) \frac{ \rp-\ro}{\vert\rp-\ro\vert^3} \, d^2 \rp\,.
\end{equation}
Note that $\=L(\ro)$ is a symmetric dyadic whose trace equals unity \cite{xYaghjian}.
 
In the limit $\rp\to\ro$, the singularity of $\=G\host(\ro,\rp;\omega)$ is offset by the singularity of $\=G_{\rm P}(\ro,\rp;\omega)$
in the integrand
of the volume integral $\#M (\ro,\omega)$. Furthermore, this integral is not improper if it is assumed that
 the electric field satisfies the H\"older
continuity condition in $\Vi$ \cite{xFikioris,xWang}: real constants $\bar{\gamma}>0$ and $\bar{\nu}>0$ exist such that
$\vert\hat{\#u}_j\.\left[\#E(\#r,\omega)-\#E(\ro,\omega)\right]\vert\leq \bar{\gamma}\vert\#r-\ro\vert^{\bar{\nu}}$ for all $\#r\in\Vi$ and $j\in\lec x,y,z\ric$. 

Provided that $\Vi$ is sufficiently small in relation to the wavelengths in both mediums, we can set $\#E(\rp,\omega)\simeq\#E(\ro,\omega)$
for all $\rp\in\Vi$, as is commonly done \cite{xWang,xFikioris,xJRNIST} following Maxwell \cite{xMaxwell} and
has been computationally verified \cite{xWang}. Then, $\#M (\ro,\omega)$ can be factored as
$\=M(\ro,\omega)\.\#E(\ro,\omega)$ and
Eq.~(\ref{xreq5}) simplifies to
\begin{equation}
\#E(\ro,\omega)\simeq \#E_{\rm s}(\ro,\omega)+\ko^2\left[\epsi(\omega)-\epsh(\omega)\right]\left[\=M(\ro,\omega)-\frac{1}{\ko^2\epsh(\omega)}\=L(\ro) \right]
\.\#E(\ro,\omega)
\,,\quad \ro\in\Vi\,.
\label{xreq8}
\end{equation}
The concept of $\=M(\ro,\omega)$ must be credited to Lorenz \cite{xLorenz1875,xPrinkey}.

Even more commonly, with good justification for really small $\Vi$ \cite{xJRNIST,xRobertson},
$\=M(\ro,\omega)$ is ignored in favor
of $\=L(\ro)/\ko^2\epsh$ \cite{xJRNIST,xYaghjian,xVanBladel} to get
\begin{equation}
\#E(\ro,\omega)\approx
\left[\=I+\frac{\epsi(\omega)-\epsh(\omega)}{\epsh(\omega)} \=L(\ro)\right]^{-1}\.
\#E_s(\ro,\omega)
\,,\quad \ro\in\Vi\,.
\label{xreq9}
\end{equation}
 
The electrically small region $\Vi$ can be said to have acquired an electric dipole moment \cite{xJRNIST}
\begin{equation}
\#p(\omega)= {\vi}\,\epso\left[\epsi(\omega)-\epsh(\omega)\right]\#E(\ro,\omega)\,
\label{xreq11}
\end{equation}
located at $\ro$
as the source of the  field scattered into $\Vh$, with
 $\vi$  denoting the volume of $\Vi$. Substituting Eq.~(\ref{xreq9}) in Eq.~(\ref{xreq11}), we get
\begin{equation}
\label{xreq12}
\#p(\omega)=  {\vi}\,\=a(\omega)\.\#E_{\rm s}(\ro,\omega)\,,
\end{equation}
where  the polarizability density dyadic
\begin{equation}
\label{xpdd-def}
\=a(\omega)\approx\epso\left[\epsi(\omega)-\epsh(\omega)\right]\left[\=I+\frac{\epsi(\omega)-\epsh(\omega)}{\epsh(\omega)}\=L(\ro)\right]^{-1}\,.
\end{equation}

The right side of Eq.~(\ref{xpdd-def})
is definitely an approximation of  $\=a(\omega)$, even for spheres, for two reasons. The first reason  is the removal of  $\=M(\ro,\omega)$  even though
$\Vi$ cannot be  infinitesimally small. The second reason is that the H\"older continuity condition can best be applied inside the largest ellipsoid 
\cite{xMaxwell} centered at $\ro$ and inscribed in $\Vi$. The electric field phasor in the portion of $\Vi$ outside that  inscribed ellipsoid having only been approximately accounted
for,   $\ro$ must be carefully chosen \cite{xMLarxiv}. Regardless, Eq.~(\ref{xpdd-def}) is  implicitly used for cubes \cite{xLivesay,xAvelin}, rectangular parallelopipeds \cite{xCauterman},  and circular cylinders
of finite height \cite{xYaghjian}. Furthermore, the analog of Eq.~(\ref{xpdd-def}) exists for quasi-two-dimensional problems \cite{xYaghjian}.

 \section{Maxwell Garnett homogenization formalism}
 
Let us now briefly go through the Maxwell Garnett formalism in this section, no novelty being implied for the derivation.
Suppose that all space $\Vsp$ be occupied by a composite medium comprising  a
 random distribution of identical, similarly oriented, electromagnetically small inclusions surrounded by a host medium.
The inclusion medium has relative permittivity scalar ${\epsi}(\omega)$, and the host medium  
 ${\epsh}(\omega)$. The volume fraction of the composite material occupied by the inclusions is denoted by   $\finc$. With $N\inc$ denoting the number density of inclusions, $\finc/N\inc$ is the volume of
each inclusion.

   The electric field in $\Vsp$ satisfies  
  \begin{equation}
 \label{xaeq1}
 \nabla\times\left[\nabla\times\#E(\#r,\omega)\right]-\ko^2\epsh(\omega)\#E(\#r,\omega)=i\omega\muo\#J(\#r,\omega)\,,\quad\#r\in\Vsp\,,
 \end{equation}
where the electric current density phasor $\#J(\#r,\omega)=\#0$ if $\#r$ lies in a region occupied by the host medium
but $\#J(\#r,\omega)\ne\#0$ if  $\#r$ lies inside an inclusion \cite{xJRNIST,xHarr}. 
Clearly, Eq.~(\ref{xaeq1}) represents a grainy visualization of the electromagnetic field in the composite medium. If this medium were
to be homogenized, then the HCM will have
\begin{equation}
\label{xaeq2}
\#D(\#r,\omega)=\epso\epsh(\omega)\#E(\#r,\omega) + \#P_{\rm xs}(\#r,\omega) =\epso  \=\eps^{\rm eff}(\omega)\.\#E(\#r,\omega)\,,
\end{equation}
as its constitutive relation, with $\#D(\#r,\omega)$ as the electric displacement phasor, $\#P_{\rm xs}(\#r,\omega)$
as the excess polarization representing the \textit{homogenized} distribution of the inclusion medium inside the HCM,
and $\=\eps^{\rm eff}(\omega)$ as the effective relative permittivity dyadic of the HCM.  

In the Maxwell Garnett homogenization formalism, Eq.~(\ref{xaeq1}) is replaced by
  \begin{equation}
 \label{xaeq1p}
 \nabla\times\left[\nabla\times\#E(\#r,\omega)\right]-\ko^2\epsh(\omega)\#E(\#r,\omega)=i\omega\muo\#J_{\rm avg}(\#r,\omega)\,,\quad\#r\in\Vsp\,,
 \end{equation}
where $\#J_{\rm avg}(\#r,\omega)=-i\omega\#P_{\rm xs}(\#r,\omega)$ is the spatial average of $\#J(\#r,\omega)$ in some small neighborhood 
of $\#r$  called the Lorentzian cavity \cite{xAspnes}. Accordingly, the
electric field phasor in the HCM
is given by
\begin{equation}
\label{xaeq3}
\#E(\#r,\omega)=\#E_{\rm cf}(\#r,\omega)+\omega^2\muo\iiint_{\Vsp} \=G\host(\#r,\rp;\omega)\.\#P_{\rm xs}(\rp,\omega)\,d^3\rp\,,
\quad\#r\in\Vsp\,,
\end{equation}
where $\#E_{\rm cf}(\#r,\omega)$ is the complementary function.

Let the electrically small spherical region $\Ve$   centered at $\#r$ be identified as the Lorentzian cavity. Then,
 we can define
\begin{equation}
\label{xaeq4}
\#E_{\rm loc}(\#r,\omega)=
\#E_{\rm cf}(\#r,\omega)+\omega^2\muo\iiint_{\Vsp-\Ve} \=G\host(\#r,\rp;\omega)\.\#P_{xs}(\rp,\omega)\,d^3\rp
\end{equation}
as the electric field phasor present at $\#r$ \textit{if} the excess polarization were to be null-valued in $\Ve$. 
Equations~(\ref{xaeq3}) and (\ref{xaeq4}) together yield
\begin{equation}
\label{xaeq5}
\#E(\#r,\omega)=\#E_{\rm loc}(\#r,\omega)+\omega^2\muo\iiint_{\Ve} \=G\host(\#r,\rp;\omega)\.\#P_{\rm xs}(\rp,\omega)\,d^3\rp\,.
\end{equation}
The conventional assumption is that the excess polarization is uniform inside a Lorentzian cavity, so that \cite{xAspnes}
\begin{equation}
\label{xaeq6}
\#E(\#r,\omega)\simeq\#E_{\rm loc}(\#r,\omega)-\frac{1}{3\epso\epsh(\omega)}\#P_{\rm xs}(\#r,\omega)
\,
\end{equation}
emerges from Eq.~(\ref{xaeq5}). 

In the Maxwell Garnett formalism, we set $\#P_{\rm xs}= N\inc \#p$ and use Eq.~(\ref{xreq12}) with $\#E_{\rm loc}(\#r,\omega)$
fulfilling the role of $\#E_{\rm s}(\#r,\omega)$; accordingly,
\begin{equation}
\label{xaeq7}
\#P_{\rm xs}(\#r,\omega) =\finc\,\=a(\omega)\.\#E_{\rm loc}(\#r,\omega)\,.
\end{equation}
We then get
\begin{equation}
\label{xaeq8}
\#E_{\rm loc}(\#r,\omega)\simeq\left[\=I-\frac{1}{3\epso\epsh(\omega)}\finc\,\=a(\omega)\right]^{-1}\.\#E(\#r,\omega)
\end{equation}
from Eq.~(\ref{xaeq6}), followed by
\begin{equation}
\label{xaeq9}
\#P_{\rm xs}(\#r,\omega) \simeq\finc\,\=a(\omega)\.\left[\=I-\frac{1}{3\epso\epsh(\omega)}\finc\,\=a(\omega)\right]^{-1}\.\#E(\#r,\omega)
\end{equation}
from Eq.~(\ref{xaeq7}).
The Maxwell Garnett estimate  
 \begin{equation} 
 \label{xepsMG}
 \=\eps^{MG}(\omega) = {\epsh}(\omega)\, \=I +  {\frac{\finc}{\epso}}\,\=a(\omega)\.\left[\=I-\frac{1}{3\epso\epsh(\omega)}\finc\,\=a(\omega)\right]^{-1}
 \end{equation}
 of the relative permittivity dyadic  $\=\eps^{eff}(\omega)$ then emerges from Eqs.~(\ref{xaeq2}) and (\ref{xaeq9}).   
 
Note that the Maxwell Garnett formalism does not really depend on any explicit formula for $\=a(\omega)$, although 
 more than a century ago, Maxwell Garnett 
 himself \cite{xMG1904} and others \cite{xSihvola} had
 used the Mossotti--Clausius expression \cite{xPrinkey} for the polarizability scalar of an electrically small, isotropic, dielectric sphere.  Parenthetically,
 even for that simple inclusion, other expressions have also been used \cite{xPrinkey, xLorenz1875,xDungey,xMallet}.
Finally, Eq.~\r{xepsMG} for $\=\eps^{MG}$  readily generalizes to cases involving inclusions of more than one type as well as to cases involving anisotropic inclusion materials \cite{xMAEH,xLakh}.

\end{document}